\documentclass[aps,prl,twocolumn,groupedaddress]{revtex4-2}
\usepackage{graphicx}
\usepackage{amsmath}

\begin{document}


\title{Coupled heat pulse propagation in two fluid plasmas}


\author{S. Jin}
\affiliation{Princeton Plasma Physics Laboratory, Princeton University, Princeton, New Jersey 08544, USA}
\author{A. H. Reiman}
\affiliation{Princeton Plasma Physics Laboratory, Princeton University, Princeton, New Jersey 08544, USA}
\author{N. J. Fisch}
\affiliation{Princeton Plasma Physics Laboratory, Princeton University, Princeton, New Jersey 08544, USA}


\date{\today}

\begin{abstract}

Because of the large mass differences between electrons and ions, the heat diffusion in electron-ion plasmas exhibits more complex behavior than simple heat diffusion found in typical gas mixtures.  In particular, heat is diffused in two distinct, but coupled, channels. Conventional single fluid models neglect the resulting complexity, and can often inaccurately interpret the results of heat pulse experiments.  However, by recognizing the sensitivity of the electron temperature evolution to the ion diffusivity, not only can previous experiments be interpreted correctly, but informative simultaneous measurements can be made of both ion and electron heat channels.
\end{abstract}


\maketitle


\textit{Introduction.}---
Electron-ion plasmas have the unique and fundamental property that the constituent populations reach equilibrium within themselves, long before equilibrating with each other. Additionally, the two species have very different transport properties—all due to the extreme mass disparity. Heat therefore propagates through two distinct, but coupled, channels. 

This basic fact has important, but thus far unaddressed, implications for heat pulse based thermal diffusivity measurements. Such experiments have been widely conducted in magnetic confinement devices since the 1980s \cite{Hartfuss_1986,Tfr_1988,Gambier_1990,Giannone_1992,Gorini_1993,Fredrickson_1986,Cox_1993,Trier_2019,Mantica_2006, Luca_1996,Zhang_2012,Jahns_1986,Zou_2003,Song_2012,Ryter_2000,Zerbini_1999, Karbashewski_2018}. Yet, many aspects of heat transport in plasmas remain mysterious. While that alone motivates their study, from a pragmatic perspective, direct measurements of the transport coefficients provide a means of validating transport models. The predictive capability of these models is critical in reactor design. These measurements also inform on profile stiffness \cite{Ryter_2001,Leuterer_2003,Tardini_2002,Imbeaux_2001} and the reduction of heat conduction within magnetic islands \cite{Spakman_2008,Inagaki_2004}. These phenomena have important consequences for disruption avoidance \cite{Reiman_2018,Rodriguez_2019,Rodriguez_2020, Jin_2020, Nies_2020}. More recently, heat pulse based transport studies have been utilized in inertial confinement fusion (ICF) experiments as well \cite{Gregori_2004, Meezan_2020}. 

The enduring relevance of heat pulse experiments has also spawned an extensive body of theoretical work devoted to their interpretation \cite{Gentle_1988, Hossain_1987,de_Haas_1991,Hogeweij_1992,Cardozo_1990,Cardozo_1995,Jacchia_1991}. While many potential deviations from diffusive propagation (e.g. convection, density coupling, non-local transport) have been considered, the basic physics of energy exchange between species, even when included, has not been properly treated. Past works model energy loss to the ions as a generic damping term for the electrons, effectively assigning the ions the status of an inactive background sink. This picture is qualitatively incomplete. If energy exchange is sufficiently significant to warrant the inclusion of a damping term, the propagation of this lost heat through the ion channel must be considered in tandem. The electron temperature is not only affected by the degree of energy exchange with the ions, but in fact bears the signature of the ion heat transport properties as well.

Beyond the direct experimental relevance, the coupled temperature modes explored in this work are also of general academic interest. There is a long legacy of work in disparate mass gas mixtures \cite{Goldman_1967,Huck_1980,Johnson_1983,Goebel_1976}, but thus far only the sound modes have received significant attention. The existence of diffusive heat modes has hardly been explored, perhaps due to limited perceived experimental relevance. Indeed, since the largest mass ratios in disparate mass neutral gases are still orders of magnitude smaller than the proton-electron mass ratio, the time scales on which a two temperature description would apply are comparatively prohibitive. Additionally, the magnetized plasma-specific tools of electron cyclotron heating (ECH) and electron cyclotron emission (ECE) are singularly well suited for exciting and measuring heat pulses respectively. Plasmas thus provide a unique opportunity to study these coupled temperature modes. 

Here we show that coupled diffusive heat pulse propagation can dramatically affect thermal diffusivity measurements. It will be shown that the electron temperature response to a periodic source generally consists of two modes, and that the propagation of these modes depends on both the electron and ion diffusivities. This superposition of modes complicates the application of broadly used single fluid based formulas, leading to varying degrees of over or under estimation of the electron diffusivity, depending on  the strength of the coupling, the distance from the source, and the ion diffusivity. Moreover, there exists the  exciting possibility of exploiting the energy coupling with ions to simultaneously extract the ion diffusivity from the electron temperature response.

\textit{1-D Diffusive heat transport.}---
The energy transport equations for electrons and ions can be written as:
\begin{equation}
\label{eq:general}
    \frac{3}{2} \partial_t n_s T_s-\nabla \cdot ( n_s  \chi_s \cdot \nabla T_s)=\frac{3}{2}\nu n_s (T_{r}-T_{s})+P_{s}
\end{equation}
where subscript $s$ denotes either electrons or ions; subscript $r$ denotes the other species; $\chi_s$ is the heat diffusivity tensor of species $s$; and $\nu$ is the electron-ion thermal equilibration rate. $P_s$ contains whatever species specific heat sources and sinks may be present, aside from the explicitly written electron-ion equilibration term. In the interest of focusing on the treatment of energy coupling, other non-diffusive effects that have already been extensively covered in other works (e.g. density coupling, convection) are excluded. In order to ensure the validity of using fluid equations for both the electrons and ions, the dynamics of interest must be slower than the ion-ion collision time, i.e.  $\tau\ll\tau_{ii}$ where $\tau$ is the time scale set by the heat source. For experiments in which heating is modulated, $\tau$ can be easily identified with the period of heating, but, for single-pulse relaxation experiments, $\tau$ is a local, less distinct quantity. 

Further simplications arise from the anisotropy of transport relative to the magnetic field. Since $\tau_{s,\perp}\gg\tau_{s,\parallel}$, where $\tau_{s,\parallel}$ ($\tau_{s,\perp}$) is the time scale on which species $s=e,i$ equilibrates along (across) field lines, the dimensionality of the problem can be reduced. For timescales $\tau\sim\tau_{s,\parallel}$, only transport along the field lines need be considered, while for timescales $\tau\sim\tau_{s,\perp}$, the temperature will already be equilibrated along field lines and only perpendicular transport will be relevant. The perpendicular transport can typically be reduced to a 1-D problem with geometrical corrections. Such corrections will not be considered here, and a slab geometry will be used for physical clarity of solutions.

As we are interested in the temperature perturbations to an equilibrium resulting from an electron heat source, all other sources and sinks are taken to be balanced and will not be explicitly written. Finally, if the density and diffusivities are weakly inhomogenous, i.e. the length scale of the perturbations is smaller than the length scale of the background in the relevant direction, the linearized 1-D equations can be approximately written: 
\begin{equation}
\label{eq:electron}
    \partial_t \widetilde T_e-X_e \partial_x^2 \widetilde T_e=\nu (\widetilde T_i-\widetilde T_e)+P_{hp}
\end{equation}
\begin{equation}
\label{eq:ion}
    \partial_t \widetilde T_i-X_i \partial_x^2 \widetilde T_i=\nu (\widetilde T_e-\widetilde T_i)
\end{equation}
where $P_{hp}$ is the external source driving the heat pulse(s), and $X_s:=\frac{2}{3}\chi_s$ for notational convenience, with the relevant diffusivity (perpendicular or parallel) determined by the time scale orderings discussed above. These notably simple coupled transport equations capture remarkably surprising and varied heat transport phenomena.

\textit{Coupled diffusive modes.}--- Heat pulse experiments often employ a periodic heat source, usually modulated electron cyclotron heating (MECH), which has the attractive property of being highly spatially localized. Then, $P_{hp}=P_0 \delta(x)\exp(-i \omega t)$, and the electron and ion temperature responses can be written as the sum of two modes: $
    \widetilde T_{\omega,s}=A_{s,1}\exp(i k_1 |x|)+A_{s,2}\exp(i k_2 |x| )$
where $k_1, k_2$ are the two solutions to 
\begin{equation}
\label{eq:disprel}
    (\nu-i \omega+X_e k^2)(\nu-i \omega+X_i k^2)=\nu^2
\end{equation}
with positive imaginary parts. As these are diffusive modes, it is always true that $\text{Im}(k_j)\geq\text{Re}(k_j)$. Modes 1 and 2 are identified by $\text{Im}(k_1)<\text{Im}(k_2)$.

The coefficients $A_{s,j}$ ($j=1,2$) are given by:
\begin{equation}
    A_{e,j}=iP/[2 X_e k_{j}(1-\frac{\alpha_{j}}{\alpha_{k\neq j}})] \quad \: \quad A_{i,j}=\alpha_jA_{e,j}
\end{equation}
where $\alpha_{j}:=\nu-i\omega+X_e k_{j}^2 $

The identity of the modes in the decoupled treatment can be easily recovered by taking the high frequency limit ($\omega/\nu\rightarrow \infty)$, or equivalently taking $\nu\rightarrow 0 $ in Eq. (\ref{eq:disprel}):
\begin{equation}
    k_1\rightarrow\frac{1+i}{\sqrt{2}}\sqrt{\frac{\omega}{X_i}} \qquad k_2\rightarrow\frac{1+i}{\sqrt{2}}\sqrt{\frac{\omega}{X_e}}
\end{equation}
Here $k_2$ can be recognized as the uncoupled electron mode, with $k_1$ as its ion counterpart. Note that, in making this identification, it has been assumed that $\chi_i>\chi_e$. Intuitively, if the driving frequency exceeds the rate at which energy can be transferred between species, the oscillations will be effectively uncoupled, with each $k$ depending solely on the respective diffusivity. The amplitude of the oscillations in the other species appropriately vanishes ($|A_{e,1}/A_{i,1}|\rightarrow 0 \:\text{ and } \:|A_{i,2}/A_{e,2}|\rightarrow 0$). The properties of each mode $k_j(\omega)$ and $A_{e,j}/A_{i,j}$ are intrinsic, but the degree to which each mode is excited depends on the heat source. Since we are considering pure electron heating, $A_{e,1}/A_{e,2}\rightarrow 0$ as $\omega/\nu\rightarrow\infty$. 

At first glance, this is an unsurprising reproduction of the single fluid treatment---intuitively it should be expected that if the driving frequency greatly exceeds the rate at which energy can be exchanged, the heat would propagate solely through the electron channel with no interference from the ions. However, if $\chi_i>\chi_e$, which is the case for perpendicular transport, then $\text{Im}(k_2)>\text{Im}(k_1)$. Then no matter how preferentially the electron mode 2 is excited, its shorter damping length means that eventually the ion diffusivity dependent mode 1 will dominate as distance from the source is increased. Of course, how much this matters in practice will depend on the measurement range, ratio of ion and electron diffusivities, and driving frequency relative to equilibration rate.

In the low driving frequency limit ($\omega/\nu\rightarrow 0$):
\begin{equation}
\begin{aligned}
        k_1 &\rightarrow\frac{1+i}{\sqrt{2}}\sqrt{\frac{2\omega}{X_e+X_i}}\\
        k_2 &\rightarrow  \frac{(X_e^2+X_i^2)}{2\sqrt{\nu X_e X_i (X_e+X_i)^3}}\omega+i\sqrt{\frac{\nu(X_e+X_i)}{X_e X_i}}
\end{aligned}
\label{eq:lowfreq}
\end{equation}

Evidently, in this limit $k_1$ describes perfectly equilibrated electrons and ions behaving as a single fluid, and propagated with an averaged diffusivity. This behaviour can be reproduced by taking the highly collisional limit of Eq. (\ref{eq:general}), leading to a reduced single fluid equation. 

The $k_2$ mode has ions oscillating perfectly out of phase with the electrons, with amplitude reduced by the factor $X_i/X_e$. It  becomes strongly damped relative to the $k_1$ mode as $\omega/\nu\rightarrow0$. The physical necessity of mode 2 arises from species specific heating--even if collisions are rapidly bringing the two species to the same temperature, close enough to the source, this perfectly equilibrated picture must break down. Accordingly, the relative amplitude of mode 2 decreases as $\omega/\nu\rightarrow0$.

In intermediate frequency regimes, the modes cannot be described in such simple terms and there is no clear preferential excitation of either mode. It is worth reiterating, however, that the mode dominance of the temperature response is not solely determined by the degree of excitation as represented by the coefficents $A_{e,j}$, but will exponentially shift in favor of the less damped mode $k_1$ with increasing distance from the source.

\textit{Diffusivity Measurements.}---
Although there are a number of ways to analyze these experiments, most often some type of analytic formula is used to relate the measured phase ($\phi$) and amplitude ($A$) profiles to the electron thermal diffusivity. The widely cited Jacchia 1991 \cite{Jacchia_1991} gives the following expression: 
\begin{equation}
\chi_e=\frac{3\omega}{4 \phi'(A'/A)}
\label{eq:jacchia}
\end{equation}

 This expression is derived accounting for electron temperature damping, but without the coupled treatment described here. This is equivalent to employing Eq. (\ref{eq:electron}) only, but forcing $\widetilde T_i=0$. Without coupling, the product of phase and amplitude derivatives is independent of the damping coefficient. 
In contrast, another commonly used expression in the literature does not account for damping (in which case  $A'/A\rightarrow\phi'$), but has the merit of only using phase measurements which are easier to reliably measure: $ \chi_e=\frac{3\omega}{4 (\phi')^2}$. In the interest of isolating the impact of the coupled treatment in particular from that of accounting for damping at all, only the former expression will be taken as a baseline for the following analysis.

\begin{figure}[h]
    \centering
 \includegraphics[width=1\linewidth]{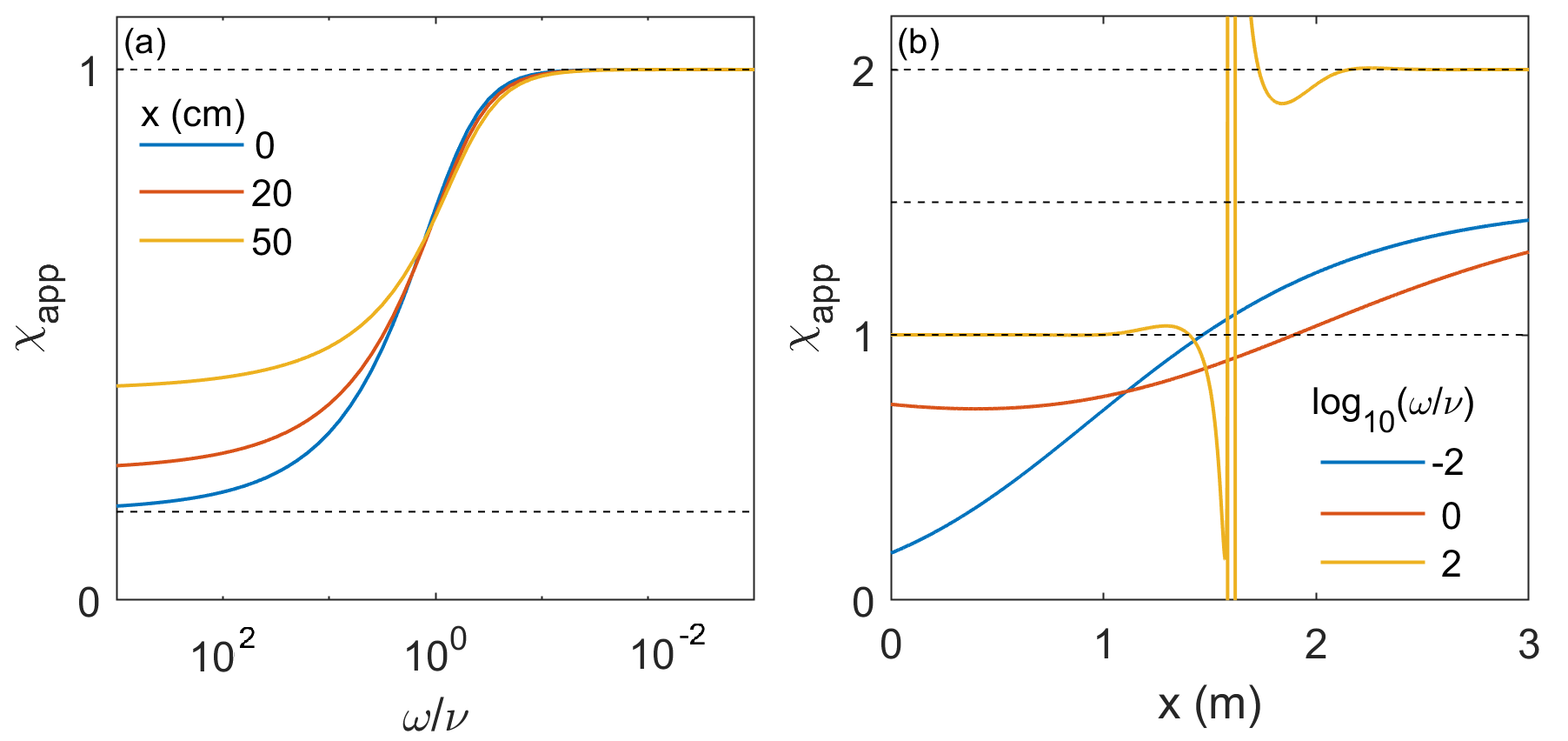}
    \caption{Apparent diffusivity $\chi_{app}$ for $\chi_e=1\; m^2/s, \chi_i=2\; m^2/s$. (a) $\chi_{app}$ vs. frequency for several distances from source. Dotted lines indicate limiting values at $\chi_e$ and $\chi_e/2(1+\gamma)$.(b) $\chi_{app}$  vs. distance from source for low, intermediate, and high driving frequencies. Dotted lines indicate limiting values at $\chi_e$, $(\chi_e+\chi_i)/2$, and $\chi_i$.}
    \label{fig:measure}.
\end{figure}

Evaluating expression (\ref{eq:jacchia}) but with phase and amplitude profiles calculated with the two fluid Eqs. (\ref{eq:electron}) and (\ref{eq:ion}) gives an ``apparent diffusivity" ($\chi_{app}$) that depends on distance from the source, driving frequency, and both electron and ion diffusivities. This quantity normalized to the true electron diffusivity is plotted in Fig. \ref{fig:measure}, which shows each of these dependencies.

Most of the limiting behaviors are obvious from the limiting forms of the solutions. At driving frequencies greatly exceeding the equilibration rate, the solution close to the source is the familiar uncoupled electron mode; in this case $\chi_{app}\rightarrow\chi_e$. Still in the high frequency limit, but far from the source, the solution is dominated by the uncoupled ion mode and accordingly $\chi_{app}\rightarrow\chi_i$.

Interestingly, at low driving frequencies, although mode 1 is both preferentially excited and not as strongly damped, the apparent diffusivity (Eq. \ref{eq:jacchia}) is not necessarily the average of the ion and electron diffusivities, $\chi_{app}\rightarrow (\chi_e+\chi_i)/2$. In fact, this is only true sufficiently far from the source. Taking derivatives of the phase or amplitude introduces factors of $k$, and as can be seen from Eq. (\ref{eq:lowfreq}), $|k_2|/|k_1|\rightarrow \infty$ as $\omega/\nu\rightarrow0$; so although the solution itself is dominated by the $k_1$ mode, the phase and amplitude derivatives contain non-negligible contributions from the $k_2$ mode. A more careful analysis yields that the true low frequency limit of expression \ref{eq:jacchia} evaluated at $x=0$ will be $\chi_e/2(1+\gamma)$, where $\gamma:=\chi_i/\chi_e$. Note that this will in general be significantly smaller than the true electron diffusivity--- a rather counter intuitive result, as one might expect the measured diffusivity to be bounded by the electron and ion values. 

\begin{figure}[h]
    \centering
 \includegraphics[width=1\linewidth]{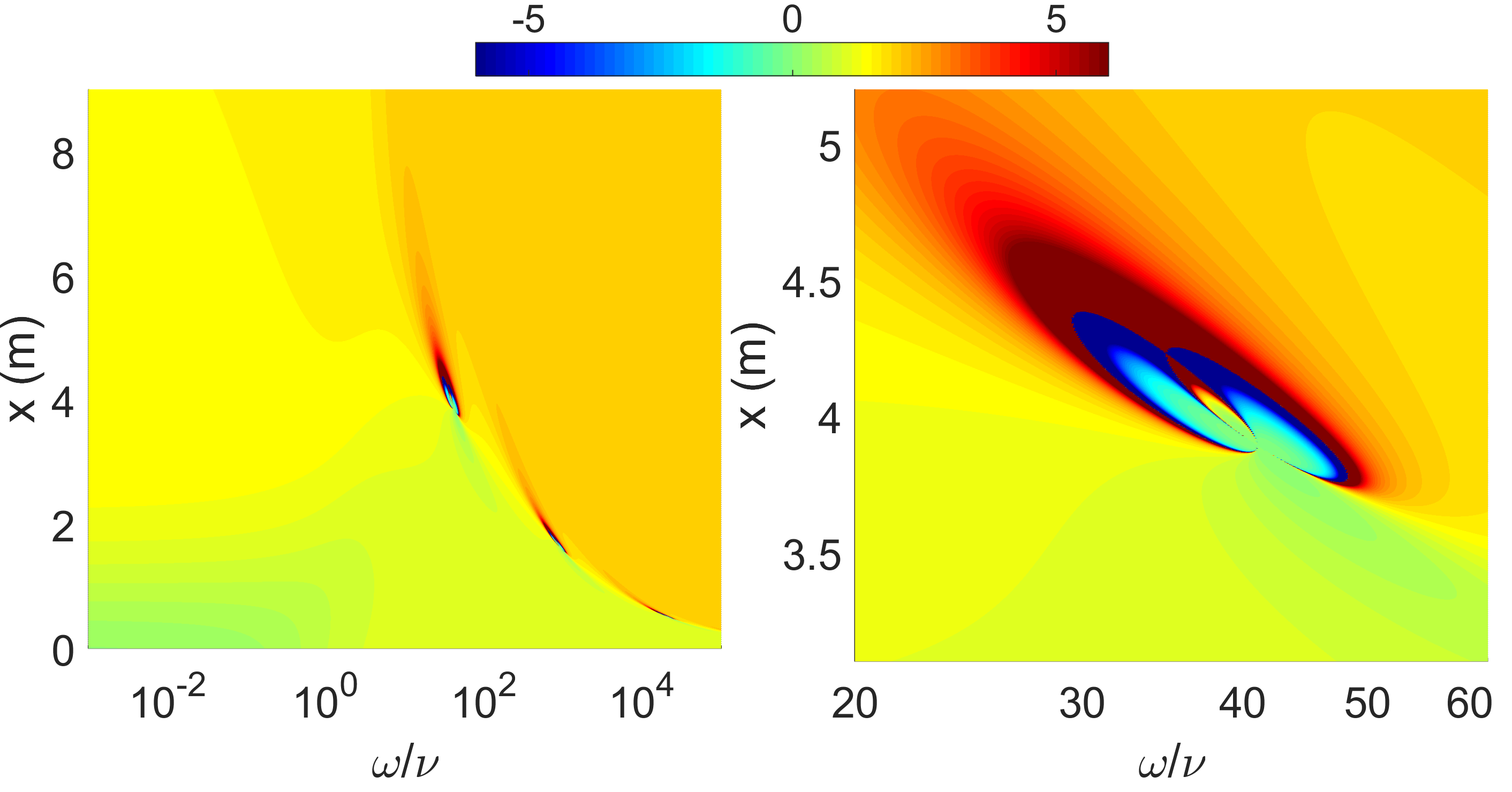}
    \caption{(Left) Relative diffusivity error $\chi_{app}/\chi_e$ as a function of measurement distance and driving frequency. (Right) Zoomed in view of a region of backwards-propagating phase and increasing amplitude. Left negative region corresponds to $\phi'<0$, right region corresponds to $A'<0$. Overlap produces positive values, so a cut along either frequency or position axis containing this overlap would exhibit 4 resonances.}
    \label{fig:surf}.
\end{figure}

At higher frequencies and larger distances from the source, Fig. \ref{fig:measure} (b) shows spiky structures in the apparent diffusivity, just before transitioning to the high frequency/far from source limits. These arise due to proximity in frequency-position space with regions of backwards-moving phase ($\phi'<0$) or increasing amplitude ($A'>0$), illustrated in Fig. \ref{fig:surf}. ``Resonances" in the apparent diffusivity appear at the boundaries of these regions, when ($\phi'=0$) or ($A'=0$). These are purely a result of the superposition of multiple modes, which in turn are only introduced by the coupling. It can be shown that Eq. \ref{eq:electron} alone without coupling does not admit backwards moving phase or locally increasing amplitude, even including, for example, inhomogeneities.

In tokamak heat pulse experiments, typically $\nu\sim10-100 \;s^{-1}$ and  $\chi_e\sim 1\; m^2/s$ so the spatial range shown in Fig. \ref{fig:measure} corresponds to $\sim 1 \; m$, while measurements are typically performed over a range on the order of $10 \;cm$. Driving frequencies are also traditionally $\mathcal{O}(10-100) \; Hz$, so while the resonances, and the ensuing far-from-source limiting behaviours are likely not of immediate experimental relevance, they cannot be ruled out. The minimalistic equations used here expose basic physical features and can provide rough estimates, but more sophisticated modeling must be performed for truly quantitiative predictions. 

The vast majority of heat pulse experiments have fallen in the high frequency/close to source regime, where effects from ion coupling are negligible. Occasionally, however, a driving frequency on the order of the equilibration rate \cite{Gambier_1990,Fredrickson_1986,Trier_2019,Ryter_2000,Ryter_2001,Tardini_2002} or even slower \cite{Zhang_2012,Zou_2003,Song_2012} is used. In this case, the apparent diffusivity can be significantly different (usually smaller) from the true electron diffusivity. These errors can be comparable with the estimated diffusivity value itself. This is in direct contrast with the consensus that neglecting ion coupling would only lead to overestimating the electron diffusivity \cite{Cardozo_1995,Goedheer_1986}.

But of far more consequence than inaccurate estimates is the missed opportunity of simultaneously extracting the ion diffusivity from the exact same electron temperature measurements that have been taken all along. Historically, higher modulation frequencies have been used to effectively drown out non-diffusive effects, ion energy exchange included, for ease of interpretation. Many experiments have strayed from this prescription, opting for slower modulation which produces oscillations that travel further for a given heating intensity, providing a better signal to noise ratio while avoiding nonlinear effects. At these lower modulation frequencies, coupling plays a significant role in the electron temperature response, and this can be exploited to infer the ion diffusivity.

\begin{figure}[h]
    \centering
 \includegraphics[width=1\linewidth]{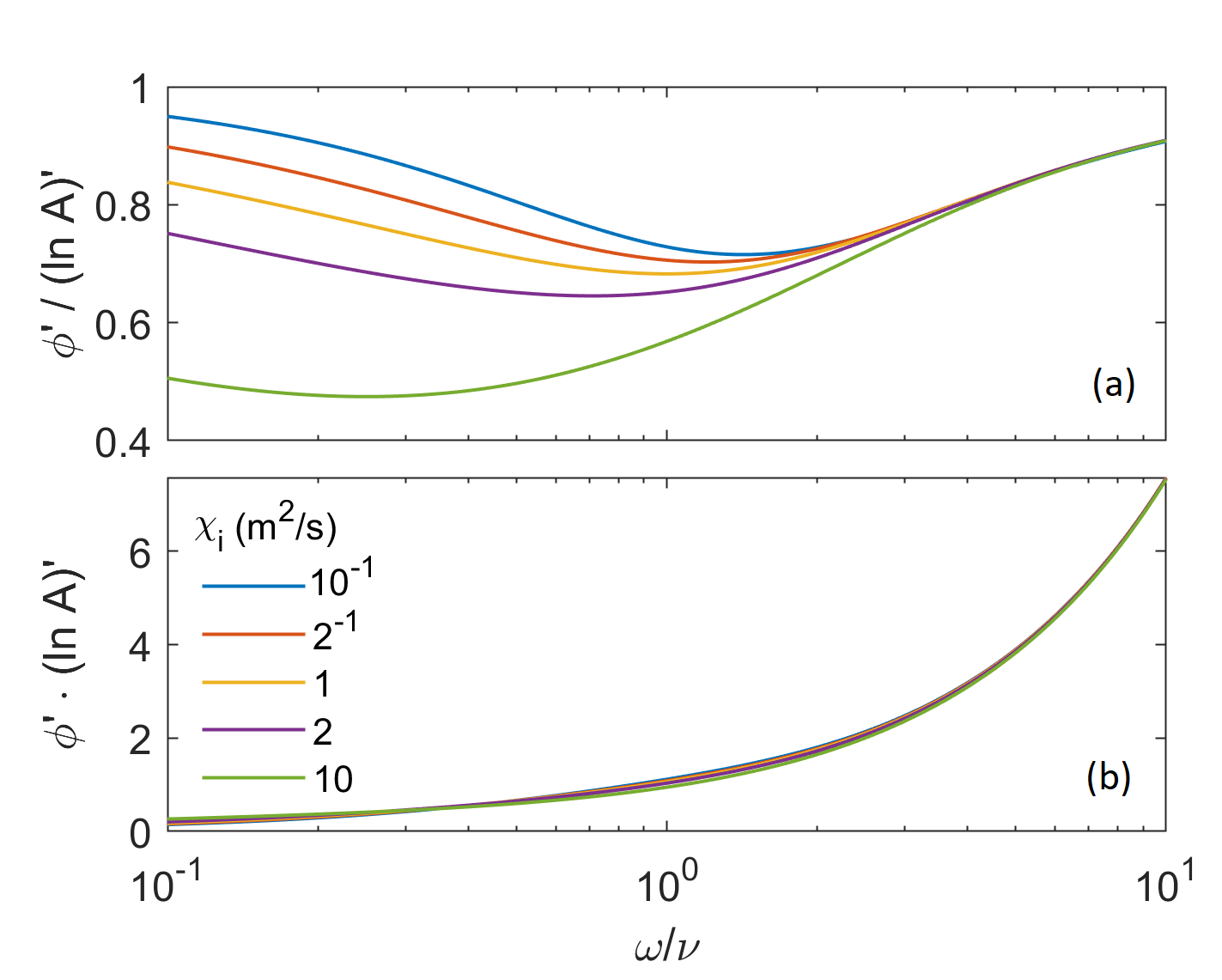}
    \caption{Composite quantites (a) $\phi'/(A'/A)$ and  (b) $\phi'\cdot(A'/A)$  evaluated at a distance of 20 cm from the source vs. driving frequency, for several ion diffusivities. }
    \label{fig:chii}.
\end{figure}

Figure \ref{fig:chii} shows the ion diffusivity and driving frequency dependence of two possible composite experimental quantities, (b) showing the combination $\phi'\cdot(A'/A)$ which is designed to be independent of damping \cite{Jacchia_1991}, (a) showing the quantity $\phi'/(A'/A)$, which was picked with the opposite intention. It can be seen that composite quantity (a) displays promising sensitivity to the ion diffusivity for driving frequencies comparable to, or slower than the local electron-ion equilibration time. This is in stark contrast to quantity (b), for which no frequency range exhibits appreciable sensitivity to the ion diffusivity, true to its design. An interesting consequence is that the apparent diffusivity evaluated with quantity (a) can be significantly different from the true electron diffusivity, while remaining insensitive to the ion diffusivity. This suggests an analysis method where both quantities are used to analyze a frequency scan---quantity (b) used to establish the local damping time and electron diffusivity, which can then be used to back out the ion diffusivity from the appropriately sensitive quantity (a).

\textit{Summary.}---
Energy exchange between species is a fundamental aspect of heat propagation in plasmas. Heat flows through two distinct, but inevitably interacting, channels with often vastly different transport properties---this can lead to surprising phenomena that can only be understood through the lens of coupled transport. The two fluid coupled treatment performed here reveals that the temperature response to a periodic source consists of two modes, a fact of far reaching consequence for diffusivity measurements.

Diffusivity values inferred with formulas derived from single fluid models may suffer from varying degrees of over or under estimation, depending not only on distance from the source and strength of coupling, but the ion diffusivity as well. A rich diversity of regimes is possible, but in typical experiments, it is most common that the electron diffusivity will be under-estimated. 

What is remarkable is that even minimal electron-ion energy coupling can change the character of heat propagation profoundly, with the electron temperature response exhibiting sensitivity to the ion heat diffusivity. This sensitivity suggests a new measurement technique: rather than attempting to avoid the energy coupling pollution of electron diffusivity measurements, it should be exploited to obtain a simultaneous estimate of the ion diffusivity. This can be accomplished simply by recognizing the coupled nature of the heat propagation in the experimental interpretation. Moreover, since the coupled heat propagation equations are linear, the spectral components of an arbitrary heat source can be treated separately by the analysis here.

As the distinct but coupled nature of heat propagation is a direct consequence of the extreme electron-ion mass disparity, the considerations here are by no means limited to magnetized plasma. Thus, heat pulse modulation techniques, well developed in the tokamak literature, but newly informed by the coupled treatment presented here, might also be adapted to other plasma settings, including inertial confinement fusion.

The authors would like to thank E. J. Kolmes for helpful discussions. This work was supported by the U.S. DOE grants DE-AC02- 09CH11466 and DE-SC0016072.
	\bibliography{bibliography}

\end{document}